\documentclass[a4paper,10pt]{article}

\usepackage[a4paper]{geometry}
\usepackage{color}
\usepackage{graphicx}
\usepackage{amsmath}
\usepackage{amsfonts}
\usepackage{amsthm}
\usepackage{amscd}
\usepackage{epsfig}
\usepackage{amssymb}
\usepackage{slashed}
\usepackage{tabularx}
\usepackage{calligra}
\usepackage{enumerate}
\usepackage{hyperref}
\usepackage{float}
\usepackage{stackrel}
\usepackage[T1]{fontenc}
\usepackage{longtable}
\usepackage{amsthm}
\usepackage{mathrsfs}
\usepackage{verbatim} 
\usepackage{fancyhdr}
\usepackage{tikz}
\usepackage{tikz-cd}
\usepackage{slashed}
\usetikzlibrary{matrix}
\usetikzlibrary{positioning}
\usepackage[all]{xy}
\usepackage{units}
\usepackage{textcomp}
\usepackage{turnstile}
\usepackage{vmargin}
\usepackage{anysize}
\usepackage{ stmaryrd }
\usepackage{tikz-3dplot}

\setmargins{2,5cm}       
{1,5cm}                        
{16cm}                      
{23,42cm}                    
{10pt}                           
{1cm}                           
{0pt}                             
{2cm}                           

\theoremstyle{plain}

\def\d{{\rm d}}
\def\i{{\rm i}}

\def\CM{\mathbb{CM}}
\def\CP{\mathbb{CP}}
\def\PT{\mathbb{PT}}

\def\tc{t_{\rm c}}
\def\xc{x_{\rm c}}
\def\yc{y_{\rm c}}
\def\zc{z_{\rm c}}
\def\uc{u_{\rm c}}
\def\vc{v_{\rm c}}
\def\wc{w_{\rm c}}
\def\wtc{\tilde{w}_{\rm c}}
\def\rc{r_{\rm c}}
\def\A{\mathcal{A}}

\def\Q{\mathcal{Q}}
\def\O{\mathcal{O}}

\begin{document}

\title{Twistor quadrics and black holes}

\author{Bernardo Araneda\footnote{Email: \texttt{bernardo.araneda@aei.mpg.de}}  \\
Max-Planck-Institut f\"ur Gravitationsphysik \\ 
(Albert-Einstein-Institut), Am M\"uhlenberg 1, \\
D-14476 Potsdam, Germany}

\date{\today}

\maketitle

\begin{abstract}
A simple procedure is given to construct curved, non-self-dual 
(complexified) K\"ahler metrics on space-time in terms of deformations 
of holomorphic quadric surfaces in flat twistor space. 
Imposing Lorentzian reality conditions, 
the Schwarzschild, Kerr, and Pleba\'nski-Demia\'nski space-times 
(among others) are derived as examples of the construction.
\end{abstract}

\section{Introduction}

Twistor theory constitutes a remarkable approach to the description of 
the complex structure of space-time 
\cite{Penrose1967, Penrose76, PMC73, Atiyah}. 
Motivated by the fact that the celestial sphere of any event in 
Minkowski space has a natural complex structure and is thus a 
Riemann sphere $\CP^1$ (cf. \cite{PR1}), the space-time manifold 
emerges as the moduli space of $\CP^1$'s (`twistor lines') 
in twistor space.
These lines are holomorphic, which implies that our intuition about lines 
and their intersection in $\mathbb{R}^{3}$ continues to hold in 
twistor space, and allows to define the conformal structure of space-time
by intersection of lines.
Gravitation, in the twistor view, should correspond to deformations 
of the flat twistor structure.
Penrose's non-linear graviton construction \cite{Penrose76} shows that 
this holds true for self-dual (or half-flat) curved space-times.

If a geometry admits a twistor space and, in addition, it has some 
further special structure, this extra structure is often holomorphically 
encoded in the twistor space. 
For example, in Riemannian geometry, Pontecorvo showed 
\cite{Pontecorvo} that a K\"ahler metric in the conformal structure of a 
conformally half-flat 4-manifold corresponds to a holomorphic section of 
(the square root of) the anti-canonical bundle of the twistor space, 
with two zeros on each twistor line.
This can also be understood in terms of a holomorphic surface 
that intersects each twistor line at two points.

Self-dual curvature is a strong restriction for general relativity, 
as it implies (conformally) flat space-time.
In this note, inspired by Pontecorvo's construction and by the non-linear 
graviton, we describe a variation of the twistor construction that 
produces non-self-dual (complexified) K\"ahler metrics on space-time 
as deformations of holomorphic quadric surfaces in flat twistor space.
Part of the basic intuition is that two points define a line; this 
carries over to twistor space since everything is holomorphic 
(the two points in question being the intersection points 
of twistor lines with the quadric; see below).
Imposing Lorentzian reality conditions, we show that the general 
Pleba\'nski-Demia\'nski class of space-times \cite{PD, Podolsky}, 
which includes the standard black hole metrics of general relativity 
such as Schwarzschild, Kerr, etc., is recovered by this construction.

\section{Twistor quadrics and K\"ahler metrics}

\subsection{Preliminaries}

We start off by introducing some basic definitions to set our conventions.
Let $(M,g)$ be a 4-dimensional, orientable Riemannian or 
Lorentzian manifold. We also allow complexified geometries.
We say that a $(1,1)$ tensor $J$ is an almost-complex structure 
if it satisfies $J^{2}=-\mathbb{I}$ and its $(\pm\i)$-eigenspaces 
$T^{\pm}$ have the same rank.
We say that $J$ is, in addition, compatible with $g$ if
it holds $g(J\cdot,J\cdot)=g(\cdot,\cdot)$; in this case 
we refer to $(g,J)$ as an almost-Hermitian structure.
The $(\pm\i)$-eigenspaces $T^{\pm}$ of $J$ split the tangent bundle
as $T^{+}\oplus T^{-}$. We say that $J$ is integrable, and is thus 
a complex structure, if $T^{\pm}$ are both involutive under the Lie 
bracket, i.e. $[T^{\pm},T^{\pm}]\subset T^{\pm}$ (for both signs). 
If this is satisfied and $J$ is also compatible with $g$, then $(g,J)$ is 
a Hermitian structure.

Given an almost-Hermitian structure, the fundamental 2-form is defined 
by $\kappa(\cdot,\cdot)=g(J\cdot,\cdot)$. 
The 2-form $\kappa$ and $J$ are compatible in the sense that
$\kappa(J\cdot,J\cdot)=\kappa(\cdot,\cdot)$.
Note that any two of $g,J,\kappa$ determines the third; in particular, 
given compatible $\kappa,J$, the metric $g$ is determined by 
$g(\cdot,\cdot)=\kappa(\cdot,J\cdot)$.
Finally, we say that the geometry is (complexified) K\"ahler if $(g,J)$ is 
Hermitian and $\d\kappa=0$. In this case, $\kappa$ is also referred to 
as the symplectic form.

Concerning reality conditions, one can show that given an 
almost-Hermitian structure, the fundamental 2-form is an eigenform 
of the Hodge star operator, so it is self-dual (SD) or anti-self-dual (ASD).
In Riemann signature, (A)SD 2-forms are real, whereas in Lorentz 
signature they are complex. 
Therefore, the tensor $J$ is real-valued in Riemann signature, 
and complex-valued in Lorentz signature. 
Lorentzian K\"ahler geometry was thoroughly investigated by 
Flaherty \cite{Flaherty}.

\subsection{Twistor space}

Let $\CM$ denote complexified Minkowski space, with complexified  
inertial coordinates $\tc,\xc,\yc,\zc$ and flat holomorphic metric 
$\eta=\d \tc^2-\d \xc^2-\d \yc^2-\d \zc^2$. 
Introduce double null coordinates $\uc,\vc,\wc,\wtc$ by 
$\uc=\frac{1}{\sqrt{2}}(\tc+\zc)$,
$\vc=\frac{1}{\sqrt{2}}(\tc-\zc)$,
$\wc=\frac{1}{\sqrt{2}}(\xc+\i\yc)$,
$\wtc=\frac{1}{\sqrt{2}}(\xc-\i\yc)$.
Twistor space is the manifold $\PT=\CP^3\backslash\CP^1$, 
and it is related to space-time via the incidence relation
\begin{align}
\left(\begin{matrix} Z^{0} \\ Z^{1} \end{matrix} \right) 
 =\i\left(\begin{matrix} 
 \uc & \wc \\ \wtc & \vc \end{matrix} \right)
 \left(\begin{matrix} Z^{2} \\ Z^{3} \end{matrix} \right)
 \label{IR}
\end{align}
where we use homogeneous coordinates $Z^{\alpha}
=(Z^{0},Z^{1},Z^{2},Z^{3})$ on $\CP^3$
(the $\CP^1$ removed corresponding to $Z^2=Z^3=0$).
The twistor correspondence \eqref{IR} is non-local.
Fixing $Z^{\alpha}\in\PT$, the set of 
space-time points satisfying \eqref{IR} is a totally null 2-surface 
in $\CM$, called `$\alpha$-surface'.
Fixing $(\uc,\vc,\wc,\wtc)\in\CM$, the set of $Z^{\alpha}$ satisfying 
\eqref{IR} is a (holomorphic, linearly embedded)
Riemann sphere $L_{x}\cong\CP^1$, which is called `twistor line'. 
Space-time is the moduli space of twistor lines in $\PT$, 
and twistor space is the moduli space of $\alpha$-surfaces in $\CM$.
Inhomogeneous local coordinates for twistor space are given by 
the equations that define $\alpha$-surfaces, that is 
(putting $\zeta=Z^3/Z^2$, in a region with $Z^2\neq0$)
\begin{align}
\omega^{0}=\uc+\zeta\wc, \qquad \omega^{1}=\wtc+\zeta\vc, 
\qquad \zeta.
\label{ICPT}
\end{align}
For fixed $(\uc,\vc,\wc,\wtc)$ and variable $\zeta$, 
these three quantities describe a twistor line in $\PT$. 
For fixed $(\omega^0,\omega^1,\zeta)$ and variable 
$(\uc,\vc,\wc,\wtc)$, 
\eqref{ICPT} describe an $\alpha$-surface in $\CM$.

Twistor space is fibered over $\CP^1$, being
the total space of the fiber bundle $\O(1)\oplus\O(1)\to\CP^1$.
Here, $\O(-1)$ is the tautological line bundle over $\CP^1$, 
and $\O(k)=\O(-1)^{*\otimes k}$ ($k>0$). 
The base of the fibration has homogeneous coordinates 
$Z^2,Z^3$, or inhomogeneous coordinate $\zeta$, 
and the fibers have coordinates $Z^0,Z^1$. 
Each fiber has an $\O(2)$-valued symplectic structure 
$\mu=\d Z^0\wedge\d Z^1$.
All of this is valid for flat space-times.
One of the main ideas in twistor theory is that gravitation, namely 
curved space-times, should 
correspond to deformations of twistor structures. 
Penrose showed \cite{Penrose76} that this is true for half-flat 
space-times: he proved that an ASD, Ricci-flat, complex space-time 
corresponds to a 3-dimensional complex manifold $\mathcal{PT}$ 
obtained as a deformation of $\PT$ that preserves the fibration 
$\mathcal{PT}\to\CP^1$ and the fiberwise symplectic structure $\mu$.

\subsection{Quadrics}

In a Riemannian setting, twistor space can also be defined as the 
space of almost-complex structures compatible with a 4-dimensional 
Riemannian conformal structure. 
This space coincides with the (6-real-dimensional) projective spin 
bundle, which can be shown to be a complex 3-manifold 
(the twistor space $\mathcal{PT}$) 
if and only if the conformal structure is ASD.

In this context, one may ask what a K\"ahler metric in the conformal 
structure corresponds to in twistor space.
This was studied by Pontecorvo \cite{Pontecorvo}, who showed 
that a (necessarily scalar-flat) K\"ahler metric corresponds to 
a preferred (global, holomorphic) section of $K^{-1/2}_{\mathcal{PT}}$ 
which vanishes at two points in each twistor line,
where $K^{-1/2}_{\mathcal{PT}}$ is the square-root of the anti-canonical 
line bundle of $\mathcal{PT}$.
We can see this by first noticing that the bundle 
$K^{-1/2}_{\mathcal{PT}}$ restricted to a twistor line is $\O(2)$. 
Hitchin showed in \cite[Section 2]{Hitchin}
that the space $H^{0}(\mathcal{PT},\O(k))$ of global, holomorphic 
sections of $\O(k)$ can be identified with $\ker(T_{k})$, where 
$T_{k}$ is the valence-$k$ twistor operator. 
This implies that a section $\chi$ of $K^{-1/2}_{\mathcal{PT}}$ 
corresponds to a valence-2 twistor-spinor, or Killing spinor. 
The requirement that $\chi$ vanishes at two points in each twistor line 
means that the Killing spinor is non-degenerate. 
Using then \cite[Lemma 2.1]{DunajskiTod}, this corresponds 
to a conformal K\"ahler structure on space-time.

In complexified flat space-time, we can also describe Pontecorvo's
construction in terms of holomorphic quadric surfaces in flat twistor 
space $\PT$.
The space $H^{0}(\CP^1,\O(k))$ consists of degree $k$ 
homogeneous polynomials in $\mathbb{C}^{2}$.
Thus, a section $\chi$ of $K^{-1/2}_{\PT}$, when restricted to 
a twistor line, is of the form 
\begin{align}
 \chi = A\zeta^2+2B\zeta+C \label{QuadraticFunction}
\end{align}
for some $A,B,C$, where we are using an inhomogeneous coordinate 
$\zeta$ on $\CP^1$. 
We can then think of $\chi$ as a holomorphic quadratic function 
$\chi(Z^{\alpha})=Q_{\alpha\beta}Z^{\alpha}Z^{\beta}$ for some 
symmetric $Q_{\alpha\beta}$. 
The expression \eqref{QuadraticFunction} follows after using the 
incidence relation \eqref{IR}, which also shows that 
$A=A(\vc,\wc)$, $B=B(\uc,\vc,\wc,\wtc)$ and $C=C(\uc,\wtc)$.
The zero set of $\chi$ is a holomorphic 
quadric, $\mathbb{Q}=\{Z^{\alpha}\in\PT \:|\: \chi(Z^{\alpha})=0\}$. 
From Kerr's theorem \cite{Penrose1967}, the surface $\mathbb{Q}$ corresponds to a shear-free, null geodesic congruence in $\CM$.
The condition that $\chi$ vanishes at two points in each twistor line 
$L_{x}$ is the same as saying that $L_x$ intersects $\mathbb{Q}$ 
at two points, corresponding to the two roots $\zeta_{\pm}$ of the 
quadratic polynomial \eqref{QuadraticFunction}, that is
$\chi=A(\zeta-\zeta_{+})(\zeta-\zeta_{-})$, where
$\zeta_{\pm}=(-B\pm\sqrt{B^2-AC})/A$.
We allow, however, the possibility of twistor lines where the roots 
coincide; these correspond to caustics in the ray congruence on 
space-time 
(and will later correspond to curvature singularities for the curved, 
non-self-dual metric we will construct). 
Importantly,  the quadric is divided into two regions in twistor space:
\begin{align}
\mathbb{Q}=\mathbb{A}_{+}\cup\mathbb{A}_{-}, \label{Apm}
\end{align}
where $\mathbb{A}_{\pm}$ can be described in 
local coordinates by any two of (see \eqref{ICPT}) 
\begin{align}
\omega^{0}_{\pm}=\uc+\zeta_{\pm}\wc, \qquad
\omega^{1}_{\pm}=\wtc+\zeta_{\pm}\vc, \qquad
\zeta_{\pm},
\label{BasicCoordQ}
\end{align}
or by any function of them. For fixed $+$ or $-$, the three coordinates in 
\eqref{BasicCoordQ} are functionally dependent as a consequence of 
the quadric equation 
$\chi(\omega^{0}_{\pm},\omega^{1}_{\pm},\zeta_{\pm})=0$.
A simple example to illustrate this (and to have in mind in general)
is a product of planes, that is, a quadric given by 
$Q_{\alpha\beta}=A^{+}_{(\alpha}A^{-}_{\beta)}$ for some fixed 
$A^{\pm}_{\alpha}$. The two regions 
in \eqref{Apm} are in this case two planes 
$\mathbb{A}_{\pm}=\{Z^{\alpha} | A^{\pm}_{\alpha}Z^{\alpha}=0 \}$, 
and the roots coincide in the twistor line corresponding to the 
intersection of the planes, see e.g. \cite[Fig. 6-11]{PR2}.

Since, generically, a twistor line $L_x$ intersects $\mathbb{Q}$ at 
two points, and since two points define a unique line through them, 
the two intersection 
points can also be used to characterise the twistor line $L_x$. 
In other words, varying the line $L_x$, the intersection points serve 
as a coordinate system on space-time.
Given local holomorphic coordinates on $\mathbb{A}_{\pm}$ 
(obtained e.g. from \eqref{BasicCoordQ}),
say $z^{A}_{\pm}$ with $A=0,1$, the pair 
$(z^{A}_{+},z^{A}_{-})$ is the desired coordinate system on $\CM$.
The complex structure $J$ induced on space-time 
from the quadric $\mathbb{Q}$ can then be shown to be 
\begin{align}
 J = \i (\partial_{z^{A}_{+}}\otimes\d z^{A}_{+} 
 - \partial_{z^{A}_{-}}\otimes\d z^{A}_{-}) 
 \label{J}
\end{align}
where the Einstein summation convention is assumed.
The tensor \eqref{J} is compatible with the Minkowski metric. 
In particular, the vectors $\partial_{z^{A}_{\pm}}$ are null.
In fact, the construction so far is conformally invariant. 
The K\"ahler structure is obtained from the symplectic form, 
which can be shown to be
\begin{align}
 \kappa = \frac{\i}{(B^2-AC)^{3/2}}[A\d\uc\wedge\d\wtc 
 - B(\d\uc\wedge\d\vc+\d\wc\wedge\d\wtc)+C\d\wc\wedge\d\vc]
 \label{KahlerForm0}
\end{align}
 ($A,B,C$ are defined in \eqref{QuadraticFunction}).
One can show this using, for example, the Penrose transform for 
spin 1, with the twistor function $f(Z^{\alpha})=[\chi(Z^{\alpha})]^{-2}$.
In terms of quadric coordinates $z^{A}_{\pm}$, the symplectic form 
\eqref{KahlerForm0} is
\begin{align}
\kappa = \kappa_{A\tilde{B}}\d z^{A}_{+}\wedge\d z^{B}_{-}, \qquad 
\kappa_{A\tilde{B}}=\kappa(\partial_{z^{A}_{+}},\partial_{z^{B}_{-}}).
\label{KahlerForm1}
\end{align}
Indices $A,\tilde{B},...$ are numerical and take values $0,1$ 
(and again Einstein summation is used).
The distinction between an index `$B$' and an index `$\tilde{B}$' is 
only intended to remind that they are associated to the two 
different halves of the quadric, and in equations like 
\eqref{KahlerForm1} they are summed over as usual.

\section{Deformed quadrics and non-self-dual K\"ahler metrics}

Consider a holomorphic quadric $\mathbb{Q}$ in twistor space, 
which is arbitrary except for the assumption that, generically, 
twistor lines intersect $\mathbb{Q}$ at two points, 
so that the quadric is divided into two regions $\mathbb{A}_{\pm}$ 
as in \eqref{Apm}. 
Choose holomorphic coordinates $z^{A}_{\pm}$ on 
$\mathbb{A}_{\pm}$.
We now introduce a ``deformed'' quadric as
\begin{align}
 \Q=\A_{+}\cup\A_{-}, \label{DeformedQuadric}
\end{align}
where $\A_{+}$ and $\A_{-}$ are the level sets of the four functions 
$\dot z^{A}_{+}$ and $\dot z^{A}_{-}$ defined by 
\begin{align}
\dot z^{A}_{+} = z^{A}_{+}, \qquad
\dot z^{A}_{-} = z^{A}_{-}+f^{A}(z^{B}_{+},z^{B}_{-}) 
\label{Deformation}
\end{align}
for some functions $f^{A}$, such that 
$\d\dot z^{0}_{+}\wedge\d\dot z^{1}_{+}\wedge 
\d\dot z^{0}_{-}\wedge\d\dot z^{1}_{-} \neq 0$. 
Although one half of the quadric remains ``undeformed'', 
$\mathbb{A}_{+}=\A_{+}$, 
the other half $\mathbb{A}_{-}$ is deformed to $\A_{-}$
and $\Q$ is, in general, {\em not} inside twistor space 
(a point in $\A_{-}$ is not in twistor space, since it does not 
correspond to an $\alpha$-surface in $\CM$).
We also note that in order to get a non-trivial construction, 
the functions $f^{A}$ must depend on both $z^{A}_{+}$ and 
$z^{A}_{-}$, otherwise \eqref{Deformation} would just be 
a diffeomorphism on the quadric. 
Although our construction is inspired by the non-linear graviton, 
the sense in which \eqref{DeformedQuadric} is a deformed quadric 
does not seem to be the same as the complex-structure-deformations 
of twistor theory.

Recalling that the complex structure on space-time induced by the 
original quadric is given by \eqref{J}, 
we associate the deformed quadric to a new complex structure:
\begin{align}
 \dot{J} = \i (\partial_{\dot z^{A}_{+}}\otimes\d\dot z^{A}_{+} 
 - \partial_{\dot z^{A}_{-}}\otimes\d\dot z^{A}_{-}). 
 \label{Jdot}
\end{align}
This is an integrable almost-complex structure on the (complexified)
space-time manifold, but it is {\em not} Hermitian: 
it is not compatible with the Minkowski metric. 
In particular, unlike the undeformed quadric, 
the new vectors $\partial_{\dot z^{A}_{+}}$ are not null 
(they are linear combinations of $\partial_{z^{A}_{+}}$ and 
$\partial_{z^{A}_{-}}$).
We then interpret the deformation \eqref{Deformation} of the quadric 
as a deformation of the conformal structure on space-time: 
new conformal structures are introduced by requiring that 
their null cones contain $\partial_{\dot z^{A}_{\pm}}$. 
 
This requirement alone, however, does not fix a metric. 
In order to do this, we must ask additional conditions on the 
deformations \eqref{Deformation}. 
To this end, we choose to restrict to quadric deformations that preserve 
the symplectic structure induced on space-time, $\kappa$. 
In some sense, we can take inspiration for this restriction 
from the non-linear graviton, where the twistor deformations 
preserve the fiberwise symplectic structure (which allows to 
reconstruct the space-time metric); 
however, this is not the same since we are here dealing 
with a symplectic structure on space-time (not on twistor space).
Regardless, the symplectic-form-preserving condition allows to 
fix a metric:
\begin{align}
 g(X,Y) := \kappa(X,\dot{J}Y), \label{Newmetric}
\end{align}
for all vectors $X,Y$, where the symmetry property of this map 
follows from requiring $\kappa$ and $\dot{J}$ to be compatible, which in 
turn is the same as requiring $\partial_{\dot z^{A}_{\pm}}$ to be null. 
The metric \eqref{Newmetric} is then
\begin{align}
 g = 2 g_{A\tilde{B}}\d\dot{z}^{A}_{+}\odot\d\dot{z}^{B}_{-},
 \label{Newmetric2}
\end{align}
where $g_{A\tilde{B}}=
g(\partial_{\dot{z}^{A}_{+}},\partial_{\dot{z}^{B}_{-}})$. 
As before, all indices here are numerical, see below eq. 
\eqref{KahlerForm1} for our conventions.
A calculation shows that the deformations \eqref{Deformation} preserve 
the symplectic structure $\kappa$ if and only if the functions 
$g_{A\tilde{B}}$ and $f^{A}$ satisfy
\begin{subequations}\label{ConditionsNewMetric}
\begin{align}
 g_{A\tilde{C}}\left(\delta^{C}_{B}
 +\frac{\partial f^{C}}{\partial z^{B}_{-}}\right)
 ={}& \i\kappa_{A\tilde{B}}, \label{ComponentsNewMetric} \\
 \epsilon^{AC}g_{A\tilde{B}}\frac{\partial f^{B}}{\partial z^{C}_{+}} ={}& 0,
 \label{Condition2}
\end{align}
\end{subequations}
where the four functions $\kappa_{A\tilde{B}}$ are defined in 
\eqref{KahlerForm1}.
The functions $g_{A\tilde{B}}$ in \eqref{Newmetric2} can then 
be computed from eq. \eqref{ComponentsNewMetric} 
(by inverting the matrix inside the brackets on the left), and the 
deformation functions $f^A$ are not completely arbitrary but are 
restricted by the condition \eqref{Condition2}.

In summary: the result of this construction is a new metric 
\eqref{Newmetric2}-\eqref{ConditionsNewMetric} on space-time that is 
generically curved, non-(A)SD, and automatically (complexified) 
K\"ahler.

\section{Black holes}

Consider the holomorphic quadric $\mathbb{Q}\subset\PT$ given as 
the zero set of the following quadratic function:
\begin{align}
\chi(Z^{\alpha})=Z^{0}Z^{3}-Z^{1}Z^{2}. \label{BasicQuadric}
\end{align}
On twistor lines, this adopts the form \eqref{QuadraticFunction} 
with $A=\wc$, $B=\frac{1}{2}(\uc-\vc)$, $C=-\wtc$ (we are omitting an 
irrelevant overall factor of $\i$ coming from \eqref{IR}). 
The roots are then easily computed to be 
\begin{align}
\zeta_{\pm}=\frac{-\zc\pm \rc}{\xc+\i\yc}, \qquad 
\rc:=\sqrt{\xc^2+\yc^2+\zc^2}. \label{roots}
\end{align}
In some sense, we could say that $\zeta_{+}$ and $\zeta_{-}$ are 
``related'' by a complexified antipodal map: 
if $\tilde{\zeta}_{\pm}=(-\zc\pm\rc)/(\xc-\i\yc)$, then 
$\zeta_{+}=-1/\tilde{\zeta}_{-}$. 
Twistor lines with $\rc=0$ intersect $\mathbb{Q}$ only once; 
the K\"ahler structure on $\CM$ is not well-defined at these points.
The symplectic form \eqref{KahlerForm0} is 
\begin{align}
\kappa = \frac{\i}{\rc^3} \left[ \xc(\d\tc\wedge\d\xc+\i\d\yc\wedge\d\zc)
+\yc(\d\tc\wedge\d\yc+\i\d\zc\wedge\d\xc) 
+\zc(\d\tc\wedge\d\zc+\i\d\xc\wedge\d\yc) \right].
\label{KahlerFormBHs}
\end{align}
Recalling \eqref{Apm} and \eqref{BasicCoordQ}, we choose the 
following quadric 
coordinates $z^{0}_{\pm}, z^{1}_{\pm}$ on $\mathbb{A}_{\pm}$: 
\begin{align}
 z^{0}_{\pm} = \omega^{0}_{\pm}, \qquad 
 z^{1}_{\pm} ={}& \tfrac{\i}{\sqrt{2}}\log(\pm\zeta_{\pm}).
 \label{CoordQuadric}
\end{align}

We now impose reality conditions: we take the real Lorentzian slice 
in $\CM$ defined by
\begin{align}
 \tc=t, \quad \xc=x, \quad \yc=y, \quad \zc=z - \i a
 \label{RealSlicePD}
\end{align}
where $t,x,y,z$ are all real, and $a$ is a real parameter.
The function $\rc$ in \eqref{roots} is complex: we denote by 
$r$ its real part, so that (from the definition of $\rc$) 
we must have $\rc = r - \i a z/r$. 
Let us introduce a real coordinate system $(r,p,\varphi)$
related to Cartesian coordinates $(x,y,z)$ by
\begin{align}
 x+\i y = \sqrt{(r^2+a^2)(1-p^2)} e^{\i\varphi}, \qquad z=rp
\end{align}
(where we assume $p^2<1$).
The symplectic form \eqref{KahlerFormBHs} 
and the quadric coordinates \eqref{CoordQuadric} become
\begin{align}
 & \kappa = \frac{\i}{(r-\i ap)^2}\left[ 
 \d t\wedge\d(r-\i ap)-\d\varphi\wedge(a(1-p^2)\d r-\i(r^2+a^2)\d p) \right],
 \label{KahlerFormPD} \\
 & z^{0}_{\pm} = \tfrac{1}{\sqrt{2}}[t\pm (r-\i ap)], \qquad
 z^{1}_{\pm} = \tfrac{1}{\sqrt{2}}\left[\varphi \pm 
 \left(-\arctan(a/r) - \tfrac{\i}{2}\log\left(\tfrac{1+p}{1-p}\right) \right)\right].
 \label{QuadricCoordPD}
\end{align}
After some calculations, we find the components $\kappa_{A\tilde{B}}$ 
in \eqref{KahlerForm1} to be
$\kappa_{0\tilde{0}} = \i\rc^{-2}$, 
$\kappa_{0\tilde{1}} = 0 =\kappa_{1\tilde{0}}$, 
$\kappa_{1\tilde{1}} = -\i\rc^{-2}(r^2+a^2)(1-p^2)$, 
where $\rc^2=(r-\i ap)^2$.

Following the prescription \eqref{Deformation}, we now 
deform the quadric given by \eqref{BasicQuadric} to a new quadric
$\Q=\A_{+}\cup\A_{-}$ according to
\begin{align}
 \dot{z}^{A}_{+}=z^{A}_{+}, \qquad 
 \dot{z}^{A}_{-}=z^{A}_{-} + R^{A}(r) + P^{A}(p), 
 \label{DefPD}
\end{align}
for arbitrary functions $R^{A}(r), P^{A}(p)$, where 
$r=r(z^{B}_{+},z^{B}_{-})$ and $p=p(z^{B}_{+},z^{B}_{-})$ 
are given by inverting the relations \eqref{QuadricCoordPD}.
A calculation shows that the symplectic-form-preserving requirement 
\eqref{Condition2} reduces to
\begin{align}
 (r^2+a^2)\frac{\partial R^1}{\partial r} 
 - a\frac{\partial R^0}{\partial r} = 0, \qquad
 a(1-p^2)\frac{\partial P^1}{\partial p} 
 - \frac{\partial P^0}{\partial p} = 0
\end{align}
so the functions $R^{0},R^{1}$ and $P^{0},P^{1}$ in \eqref{DefPD} 
are not independent but are related by this condition.
The new metric on space-time is given by \eqref{Newmetric2}, 
\eqref{ComponentsNewMetric}, and, as mentioned, it is curved, 
non-(A)SD, and (complexified) K\"ahler. 
Furthermore, it turns out that this simple prescription already identifies 
the Pleba\'nski-Demia\'nski class \cite{PD, Podolsky}: 
to see this, we define four functions 
${\rm T}(t,r,p), \Phi(\varphi,r,p), \Delta_r(r),\Delta_p(p)$ by 
\begin{align}
 & {\rm T}:= t + \tfrac{1}{\sqrt{2}}(R^{0}(r)+P^{0}(p)), \qquad
 \Phi:= \varphi + \tfrac{1}{\sqrt{2}}(R^{1}(r)+P^{1}(p)), 
  \label{tphi-pd} \\
 & \Delta_{r}:=(r^2+a^2)
 \left(1-\tfrac{1}{\sqrt{2}}\tfrac{\partial R^{0}}{\partial r} \right)^{-1}, 
 \qquad
 \Delta_{p}:=\i a(1-p^2)
 \left(\i a + \tfrac{1}{\sqrt{2}}\tfrac{\partial P^{0}}{\partial p} \right)^{-1}.
 \label{Deltaspd}
\end{align}
After some lengthy calculations, the new metric \eqref{Newmetric2}, 
\eqref{ComponentsNewMetric} is 
\begin{equation}
\begin{aligned}
g = \frac{1}{\rc^2} & \left[   
 \frac{(\Delta_r-a^2\Delta_p)}{\Sigma}\d{\rm T}^2 + 
 \frac{2a[(r^2+a^2)\Delta_p-(1-p^2)\Delta_r)}{\Sigma}
 \d{\rm T}\d\Phi \right. \\
 &\left. +\frac{[a^2(1-p^2)^2\Delta_r-(r^2+a^2)^2\Delta_p]}{\Sigma}
 \d\Phi^2
-\frac{\Sigma}{\Delta_r}\d r^2-\frac{\Sigma}{\Delta_p}\d p^2\right], 
 \label{PD}
\end{aligned}
\end{equation}
where $\Sigma:=r^2+a^2p^2$. 
By choosing a specific form for $\Delta_r,\Delta_p$, this is the K\"ahler 
metric associated to the Pleba\'nski-Demia\'nski space-time \cite{AA22}.

We emphasise that the definitions \eqref{tphi-pd}-\eqref{Deltaspd} 
are introduced only to recover the familiar form \eqref{PD}: 
all necessary information about the metric \eqref{PD}
is already contained in the deformed quadric \eqref{DefPD}. 

As an example, put first $R^1=P^0=P^1=0$ and then $a=0$ 
(so that $\Delta_p$ reduces to $1-p^2$), define $\cos\theta:=p$ 
and $f(r):=\Delta_r/r^2$; then \eqref{PD} multiplied by $r^2$ is 
the (real, ordinary) Schwarzschild metric if one sets $\Delta_r=r^2-2Mr$. 
Similarly, the Reissner-N\"ordstrom metric, and cosmological versions, 
etc., are obtained by choosing different functions $\Delta_r$. 
Space-time points with $r=0$, corresponding to twistor lines intersecting 
the undeformed quadric $\mathbb{Q}\subset\PT$ only once, 
are curvature singularities.

As another example, put $P^0=P^1=0$. Defining $\cos\theta:=p$, 
and setting $\Delta_r=r^2-2Mr+a^2$, the metric \eqref{PD} 
multiplied by $\rc^2$ is the (real) Kerr metric. 
The Kerr-Newman metric corresponds to $\Delta_r=r^2-2Mr+a^2+Q^2$, 
and to obtain the cosmological versions one must include 
non-trivial $P^0,P^1$.
Twistor lines intersecting $\mathbb{Q}$ only once are those with 
$\rc=0$, which is the same as $r=0=\cos\theta$ and correspond 
to ring singularities.

\section{Final remarks}

The particular quadric \eqref{BasicQuadric} used in the derivation of 
the Pleba\'nski-Demia\'nski space-time can be given some sort 
of physical interpretation, by writing it as 
$Q_{\alpha\beta}Z^{\alpha}Z^{\beta}$ 
and noticing that $Q_{\alpha\beta}$ is the angular-momentum twistor 
corresponding to a static, spin-less particle at rest in a complex 
space-time. In fact, the Penrose transform can be used here to show 
that the associated twistor functions produce the spin 2 field of 
linearized black holes \cite{PMC73}.
This idea has been revived in recent interesting work on scattering 
amplitudes, see e.g. \cite{Guevara}. 
Our construction shows, however, that the exact non-linear solutions 
are associated to a deformation of the quadric (which is not inside 
twistor space), in line with the general twistor philosophy that curved 
space-times should correspond to deformed twistor structures.

The approach in this work has been to take a complexified space-time 
as a starting point, and then recover real slices by the imposition of 
(Lorentzian) reality conditions. This is why we needed to consider 
only one quadric \eqref{BasicQuadric} to recover different space-times. 
If, on the contrary, we assume from the beginning that $\tc,\xc,\yc,\zc$ 
in \eqref{IR} are real, then the quadrics for (say) Schwarzschild and 
Kerr are different.
Also, we have chosen to work with the form of the 
Pleba\'nski-Demia\'nski space-time given in \cite{Podolsky}, 
as this allows to recover standard black holes in a straightforward 
manner. If we wish to work with the original Pleba\'nski-Demia\'nski 
coordinates \cite{PD}, one possibility is to start from a twistor quadric 
different from \eqref{BasicQuadric}. The corresponding quadric is 
$\chi=Z^{0}Z^{1}+cZ^{2}Z^{3}$, as was found by 
Haslehurst and Penrose \cite{PH}.

There are many open questions concerning our construction 
that we believe deserve further investigation. 
Can any (Riemannian or Lorentzian) K\"ahler metric be obtained by 
this procedure? 
In particular, the Chen-Teo instanton \cite{ChenTeo}? 
Also, the Einstein equations in the non-linear graviton are automatically
encoded in the deformed twistor space; 
how are field equations encoded in the deformations considered 
in this work? 
It would also be desirable to obtain a more intrinsic (i.e. not coordinate 
dependent) approach to the deformations.
Finally, ``non-integrable'' deformations of the quadric might be related 
to black hole perturbation theory, since metric perturbations constructed 
from the Teukolsky equations still possess one family (but not two) 
of $\alpha$-surfaces \cite{Araneda19}.

\medskip
\noindent
{\em Acknowledgements.} 
I would like to thank S. Aksteiner and L. Andersson for discussions.
The author gratefully acknowledges the support of the Alexander von 
Humboldt foundation.

\end{document}